# On the radiative and thermodynamic properties of the cosmic radiations using *COBE* FIRAS instrument data: IV. Sunyaev-Zel'dovich distortion effect

Anatoliy I Fisenko, Vladimir F Lemberg

*ONCFEC Inc., 250 Lake Street, Suite 909, St. Catharines, Ontario L2R 5Z4, Canada*
*Phone: 905-931-9097, E-mail: afisenko@oncfec.com*

**Abstract** The Sunyaev-Zel'dovich effect is a small spectral distortion of the spectrum of the cosmic microwave background (CMB) radiation. This slight distortion is described by the Bose-Einstein (µ-type) distribution with non-zero chemical potential. It is now interesting to investigate the effect of this distortion on the integral characteristics of the Bose-Einstein spectrum. The thermal radiative and thermodynamic functions of the Bose-Einstein distribution are such integral characteristics. These functions are as follows: a) the total radiation power per unit area; b) total energy density; c) number density of photons; d) grand potential density; e) Helmholtz free energy density; f) entropy density; g) heat capacity at constant volume; h) enthalpy density; and i) pressure. The exact analytical expressions are obtained for the temperature dependences of these functions. Using experimental data measured by the *COBE* FIRAS instrument, the thermal radiative and thermodynamic functions are calculated at the monopole temperature $T = 2.72548$ K and at $z = 0$. A comparative analysis of the results obtained with the results for the same functions of the CMB spectrum is carried out. The thermal radiative and thermodynamic functions of the Bose-Einstein distribution are calculated in the redshift range $10^5 < z < 3 \times 10^6$. The expressions are obtained for new astrophysical parameters, such as the entropy density/Boltzmann constant and number density, created by the µ - distortion of the CMB spectrum.

## 1 Introduction

It is well-known that the Sunyaev-Zel'dovich effect (Sunyaev & Zel'dovich, 1980) is a small spectral distortion of the cosmic microwave background (CMB) spectrum caused by the scattering of the CMB photons on high energy electrons (Compton scattering). After many Compton scattering, photons and electrons reach statistical equilibrium, which leads to the use of the Bose-Einstein distribution with a non-zero chemical potential. This effect is important when we consider the epoch between $z = 10^5$ and $z = 3 \times 10^6$ (Mather & Hinshaw & Page 2013).

It is well known that in order to obtain a clear picture of the thermal history of the universe, we must have information about the thermal radiative and thermodynamic properties of cosmic radiation for each epoch. This article is one of a set of articles associated with the study of the thermal radiative and thermodynamic properties of the cosmic radiations using the *COBE* FIRAS observation data. In previous articles, these properties have been discussed in detail for the cosmic microwave background radiation (Fisenko & Lemberg 2014 a; Fisenko & Lemberg 2016), the extragalactic far-infrared background radiation (Fisenko & Lemberg 2014 b), and the galactic far-infrared radiation (Fisenko & Lemberg 2015). The exact analytical expressions for temperature and redshift dependences of the thermal radiative and thermodynamic functions of the cosmic radiations have been obtained in each cases. These functions have been calculated using the *COBE* FIRAS observation data. New astrophysical parameters, such as the entropy density/Boltzmann constant and the number density of photons have been constructed.

The main idea of this study is to obtain the analytical expressions for the integral characteristics of the Bose-Einstein (µ-type) spectrum. These integral characteristics include the thermal radiative and thermodynamic functions of a system. The latter functions are: a) the total radiation power per unit area; b) total energy density; c) number density of photons; d) grand potential density; e) Helmholtz free energy density; f) entropy density; g) heat capacity at constant volume; h) enthalpy density; and i) pressure. Using *COBE* FIRAS observations, the values of thermal radiative and thermodynamic functions of the Bose-Einstein (µ-type) spectrum are calculated at the monopole temperature $T = 2.72548$ K. A comparative analysis of the results obtained with the results for the same functions of the CMB radiation is carried out. The thermal radiative and thermodynamic functions of the Bose-Einstein (µ-type) spectrum are calculated at the redshift $z = 10^5$ and $z = 3 \times 10^6$. New astrophysical parameters, created by the µ - distortion of the CMB spectrum are constructed.

## 2 General relationships for the Bose-Einstein distribution

According to (Fixsen et al. 1996), the Planck function in the redshift range $10^5 < z < 3 \times 10^6$ at given temperature $T$ is modified by the Bose-Einstein (µ-type) distribution

$$I(v,T) = \frac{8\pi h}{c^3} \frac{v^3}{e^{\frac{hv}{k_B T}+\mu} - 1} , \qquad (1)$$

where $|\mu| < 9 \times 10^{-5}$ (95% CL) is the dimensionless chemical potential (Fixsen et al. 1996), $T$ is the temperature,

Using Eq. (1), the total energy density of the Bose-Einstein (µ-type) distribution in the frequency domain is defined as (Landau & Lifshitz 1980)

$$I_0(T) = \int_0^\infty I_v(T) dv = \frac{8\pi h}{c^3} \int_0^\infty \frac{v^3}{e^{\frac{hv}{k_B T}+\mu} - 1} dv. \qquad (2)$$

Using relationship between the total energy density and the total radiation power per unit area $I^{SB}(T) = \frac{c}{4\pi} I_0(T)$ for the µ - distortion Stefan-Boltzmann law can be determined as follows

$$I^{SB}(v_1, v_2, T) = \frac{2h}{c^2} \int_0^\infty \frac{v^3}{e^{\frac{hv}{k_B T}+\mu} - 1} dv . \qquad (3)$$

The number density of the Bose-Einstein distribution $n = \frac{N}{V}$ has the following form (Landau & Lifshitz 1980):

$$n = \frac{8\pi h}{c^3} \int_0^\infty \frac{v^2}{e^{\frac{hv}{k_B T}+\mu} - 1} dv . \qquad (4)$$

Taking into account the µ - distortion, the thermodynamic functions of CMB radiation can be presented as (Landau & Lifshitz 1980):

1) The grand potential density Ω:

$$\Omega(T) = \frac{8\pi k_B T}{c^3} \int_0^\infty v^2 \ln\left(1 - e^{\frac{hv}{k_b T} + \mu}\right) dv \quad . \tag{5}$$

Here $\Omega = I_o - Ts - \mu N$, where $I_o$, $S$, and $\mu$ are the total energy density, the entropy density, and chemical potential density of a system.

2) Helmholtz free energy density $f = \frac{F}{V}$:

$$f = \Omega - \mu\left(\frac{\partial \Omega}{\partial \mu}\right)_{T,V} \quad . \tag{6}$$

3) Entropy density $s = \frac{S}{V}$:

$$s = -\left(\frac{\partial \Omega(T)}{\partial T}\right)_V \quad . \tag{7}$$

4) Heat capacity at constant volume per unit volume $c_V = \frac{C_V}{V}$:

$$c_V = \left(\frac{\partial I_0(T)}{\partial T}\right)_V \quad . \tag{8}$$

5) Pressure $p$:

$$p = -\Omega \quad . \tag{9}$$

6) Enthalpy density $h = u + p$

$$h = -\mu\left(\frac{\partial \Omega}{\partial \mu}\right)_{T,V} - T\left(\frac{\partial \Omega}{\partial T}\right)_{\mu,V} \quad . \tag{10}$$

**3. Results for the thermal radiative and thermodynamic functions of the Bose-Einstein (μ-type) spectrum**

According to Eqs. (2-4), after computing the integrals in Eq. 2 and Eq. 4, the exact expressions for the total energy density, the total radiation power per unit area (μ- distortion Stefan-Boltzmann law), and the number density of photons have the following forms:

a) The total energy density

$$I_0(T) = \frac{48\pi(k_B T)^4}{c^3 h^3} \mathrm{Li}_4(e^{-\mu}) \quad , \tag{11}$$

where $\Gamma(x)$ is the gamma function and $\mathrm{Li}_n(x)$ is the polylogarithm function (Abramowitz & Stegun 1972).

b) The $\mu$ - distortion Stefan-Boltzmann law

$$I_0(T) = \sigma_0 T^4 . \tag{12}$$

where $\sigma_0 = \dfrac{12\pi k_B^4}{c^2 h^3} \mathrm{Li}_4(e^{-\mu})$.

c) The number density of photons

$$n = \frac{16\pi(k_B T)^3}{c^3 h^3} \mathrm{Li}_3(e^{-\mu}) . \tag{13}$$

Using Eqs. (5-10) the exact expressions for the thermodynamic functions of the Bose-Einstein (µ-type) spectrum have the following forms:

1) Grand potential density

$$\Omega = -\frac{16\pi(k_B T)^4}{c^3 h^3} \mathrm{Li}_4(e^{-\mu}) . \tag{14}$$

2) Helmholtz free energy density:

$$f = -\frac{16\pi(k_B T)^4}{c^3 h^3} \mathrm{Li}_4(e^{-\mu}) \left[1 + \frac{\mathrm{Li}_3(e^{-\mu})}{\mathrm{Li}_4(e^{-\mu})} \mu \right] . \tag{15}$$

3) Entropy density

$$s = \frac{64\pi k_B^4 T^3}{c^3 h^3} \mathrm{Li}_4(e^{-\mu}) \left[1 + \frac{\mathrm{Li}_3(e^{-\mu})}{4\mathrm{Li}_4(e^{-\mu})} \mu \right] . \tag{16}$$

4) Heat capacity at constant volume per unit volume

$$c_V = \frac{192\pi k_B^4}{c^3 h^3} T^3 \mathrm{Li}_4(e^{-\mu}) \left[1 + \frac{\mathrm{Li}_3(e^{-\mu})}{4\mathrm{Li}_4(e^{-\mu})} \mu \right] . \tag{17}$$

5) Pressure of photons

$$p = \frac{16 k_B^4}{c^3 h^3} T^4 \mathrm{Li}_4(e^{-\mu}) . \tag{18}$$

6) Enthalpy density

$$h = \frac{16k_B^4}{c^3 h^3} T^4 \text{Li}_4(e^{-\mu}) \ . \tag{19}$$

Table 1 shows the calculated values of thermal radiative and thermodynamic functions of the Bose-Einstein (µ-type) spectrum and monopole spectrum at $T = 2.72548$ K at present day $z = 0$. As can be seen from Table 1, the number density of photons created by the µ- distortion of CMB spectrum, when $\mu = 9 \times 10^{-5}$, is less than for the number density of photons of the monopole spectrum. The percentage of photon destruction is $\tilde{n} = \frac{n^m - n^\mu}{n^m} \times 100\% = 0.015\%$. But for $\mu = -9 \times 10^{-5}$, the number density of photons, created by the µ- distortion of CMB radiation is larger than for the one of the monopole spectrum. This means that photons are injected in a system. The percentage of photons injected is $|\tilde{n}'| \approx 0.01\%$. As seen, this percentage is very small. The same situation arises for other thermal radiative and thermodynamic functions.

Let us calculate the value of integral distortion for the total energy density, for example. According to Table 1, when $\mu = 9 \times 10^{-5}$, we obtain $\frac{I_0^M - I_0^\mu}{I_0^M} \approx 1.2 \times 10^{-5}$. For the entropy density, we have $\frac{s^M - s^\mu}{s^M} \approx 9.7 \times 10^{-5}$. As can be clearly seen, these values are in the same order as for the average spectral distortion, according to Cobe/Firas data $\frac{\Delta I(\nu)}{I(\nu)} \simeq 10^{-5} - 10^{-4}$.

Eqs. (11-19) describe the temperature dependences of the thermal radiative and thermodynamic properties of the Bose-Einstein (µ-type) spectrum. To convert them to the redshift z-dependency, the following relationship $T(z) = T_0(1+z)$ should be used. Then the dimensionless chemical potential can be represented in the form:

$$|\mu| < 9 \times 10^{-5} (1+z) \ . \tag{20}$$

Using Eq. (20), the dimensionless chemical potential is defined as: a) $\mu = 9$ and $\mu = -9$, when $z = 10^5$; and b) $\mu = 270$ and $\mu = -270$, when $z = 3 \times 10^6$. The calculation for the thermal radiative and thermodynamic properties of the Bose-Einstein (µ-type) spectrum was performed only for positive values of $\mu = 9$ and $\mu = 270$. The reason is follows. The polylogarithm functions $\text{Li}_4(e^{-\mu})$ and $\text{Li}_3(e^{-\mu})$ in Eqs. (11-19) for $\mu = -9$ and $\mu = -270$ are infinity. As a result, all thermal radiative and thermodynamic values are also equal to infinity.

In Table 2 and Table 3, a comparative analysis of the thermal radiative and thermodynamic functions of Bose-Einstein (µ-type) spectrum with the similar functions of the monopole spectrum at redshifts $z = 10^5$ and $z = 3 \times 10^6$ are represented. As we can see from Table 2 and 3, the integral distortion for the total energy density, for example, when $\mu = 9$ and $\mu = 270$ is $\frac{I_0^M - I_0^\mu}{I_0^M} \approx 0.999$. This mean that the CMB radiation strongly dominates in comparison with the Bose-Einstein (µ-type) radiation. The same situation is observed for other thermal radiative and thermodynamic functions.

### 4. Astrophysical constants and parameters

The expressions obtained for the thermal radiative and thermodynamic functions are important for constructing new astrophysical parameters. Indeed, according to the table of astrophysical constants and parameters (Groom 2013), two fundamental parameters for the monopole spectrum, such as the number density of CMB photons and the entropy density/Boltzmann are of interest.

Let us construct the same astrophysical parameters for the Bose-Einstein (µ-type) spectrum. According to Tables 1, 2, 3 and using Eq. 13 and Eq. 16, we obtain the following expressions for new astrophysical parameters:

1. Entropy density/Boltzmann constant

$$\frac{s}{k_B} = A \left(\frac{T}{T_0}\right)^3 \left(\frac{\text{Li}_4(e^{-\mu})}{\text{Li}_4(e^{-\mu_0})}\right) \left\{ \left[1 + \frac{\text{Li}_3(e^{-\mu})}{4\text{Li}_4(e^{-\mu})}\mu\right] \Big/ \left[1 + \frac{\text{Li}_3(e^{-\mu_0})}{4\text{Li}_4(e^{-\mu_0})}\mu_0\right] \right\} \text{ cm}^{-3}, \qquad (21)$$

where

$$A = \frac{64\pi k_B^3 T_0^3 \text{Li}_4(e^{-\mu_0})}{c^3 h^3} \left[1 + \frac{\text{Li}_3(e^{-\mu_0})}{4\text{Li}_4(e^{-\mu_0})}\mu_0\right] . \qquad (22)$$

2. Number density of the photons for CMB spectral µ-distortion

$$n = C \left(\frac{T}{T_0}\right)^3 \left(\frac{\text{Li}_3(e^{-\mu})}{\text{Li}_3(e^{-\mu_0})}\right) \text{cm}^{-3}, \qquad (23)$$

where

$$C = \frac{16\pi (k_B T_0)^3 \text{Li}_3(e^{-\mu_0})}{c^3 h^3}. \tag{24}$$

Here $T_0$ = 2.72548 K and $|\mu_0| < 9 \times 10^{-5}$ are the values of the temperature and the dimensionless chemical potential at present day. $T(z)$ and $\mu(z)$ are the variable values in the redshift range from $z = 10^5$ to $z = 3 \times 10^6$.

## 3  Conclusions

In this paper, the exact expressions for the calculation of the temperature dependences of the thermal radiative and thermodynamic functions of the Bose-Einstein spectrum are obtained. Such functions are: a) the total radiation power per unit area; b) total energy density; c) number density of photons; d) grand potential density; e) Helmholtz free energy density; f) entropy density; g) heat capacity at constant volume; h) enthalpy density; and i) pressure.

Utilizing experimental data measured by the *COBE* FIRAS instrument the thermal radiative and thermodynamic functions of the Bose-Einstein spectrum are calculated at present day at the monopole temperature $T$ = 2.72548 K and $|\mu| < 9 \times 10^{-5}$ (95% CL). The results are presented in Table 1. A comparative analysis of the results obtained with the results of the monopole spectrum of the CMB radiation is performed. Small distortions for the thermal radiative and thermodynamic functions are observed. For example, for number density of photons, we obtain: $\frac{n^M - n^\mu}{n^\mu} \simeq 1.5 \times 10^{-4}$ when $\mu = 9 \times 10^{-5}$.

Knowing the dependence of the temperature $T$ on the redshift z allows us to study the state of the Universe many years ago. Therefore, the thermal radiative and thermodynamic functions of the Bose-Einstein spectrum are calculated at the redshifts z = $10^5$ and z = $3 \times 10^6$. Table 2 and Table 3 represent the calculated values. A comparative analysis of the results obtained with the results for the monopole spectrum at $T = 2.72550 \times 10^5$ K and $T = 8.176442 \times 10^6$ K is presented. It is shown that the CMB radiation strongly dominates in comparison with the Bose-Einstein (μ-type) radiation.

The analytical expressions for the thermal radiative and thermodynamic functions for the Bose-Einstein spectrum allow us to construct new astrophysical parameters, such as the entropy density/Boltzmann constant, and number density of $\mu$- distortion photons.

In conclusion, it is important to note the following. It is well-know that in order to construct cosmological models of heat transfer the Stephan-Boltzmann law in the form $\sigma T^4$ is typically used. If we consider an epoch from $z = 10^5$ to $z = 3 \times 10^6$, we must also add to this law the contribution of the $\mu$- distortion component Eq. (13). This fact should be taken into account when we consider cosmological models for calculation.

**Acknowledgments**

The authors sincerely thank Professor A. Zhuk and Professor N.P. Malomuzh for fruitful discussions.

| Quantity | $\mu$-distortion spectrum $\mu = 9 \times 10^{-5}$ | $\mu$-distortion spectrum $\mu = -9 \times 10^{-5}$ | Monopole spectrum $\mu = 0$ |
|---|---|---|---|
| $I_0(T)$ $[\text{J m}^{-3}]$ | $4.1742 \times 10^{-14}$ | $4.1751 \times 10^{-14}$ | $4.1747 \times 10^{-14}$ |
| $I_0^{SB}(T)$ $[\text{W m}^{-2}]$ | $3.1285 \times 10^{-6}$ | $3.1291 \times 10^{-6}$ | $3.1288 \times 10^{-6}$ |
| $\Omega(T)$ $[\text{J m}^{-3}]$ | $-1.3914 \times 10^{-14}$ | $-1.3917 \times 10^{-14}$ | - |
| $f$ $[\text{J m}^{-3}]$ | $-1.3915 \times 10^{-14}$ | $-1.3915 \times 10^{-14}$ | $-1.3915 \times 10^{-14}$ |
| $s$ $[\text{J m}^{-3}\text{ K}^{-1}]$ | $2.0421 \times 10^{-14}$ | $2.0424 \times 10^{-14}$ | $2.0423 \times 10^{-14}$ |
| $P$ $[\text{J m}^{-3}]$ | $1.3914 \times 10^{-14}$ | $1.3917 \times 10^{-14}$ | $1.3915 \times 10^{-14}$ |
| $c_V$ $[\text{J m}^{-3}\text{ K}^{-1}]$ | $6.1264 \times 10^{-14}$ | $6.1273 \times 10^{-14}$ | $6.1269 \times 10^{-14}$ |
| $n$ $[\text{m}^{-3}]$ | $4.1066 \times 10^{8}$ | $4.1076 \times 10^{8}$ | $4.1072 \times 10^{8}$ |
| $h(T)$ $[\text{J m}^{-3}]$ | $1.3914 \times 10^{-14}$ | $1.3917 \times 10^{-14}$ | $1.3916 \times 10^{-14}$ |

**Table 1** Calculated values of the radiative and thermodynamic functions for Bose-Einstein ($\mu$-type) spectrum and monopole spectrum at present day $z = 0$ and $T = 2.72548$ K.

| Quantity | $\mu$-distortion spectrum $\mu = 9$ | Monopole spectrum $\mu = 0$ |
|---|---|---|
| $I_0(T)$ [J m$^{-3}$] | $4.7603 \times 10^2$ | $4.1748 \times 10^6$ |
| $I_0^{SB}(T)$ [W m$^{-2}$] | $3.5677 \times 10^{10}$ | $3.1289 \times 10^{14}$ |
| $\Omega(T)$ [J m$^{-3}$] | $-1.5867 \times 10^2$ | - |
| $f$ [J m$^{-3}$] | $-1.5867 \times 10^3$ | $-1.3916 \times 10^6$ |
| $s$ [J m$^{-3}$ K$^{-1}$] | $7.5685 \times 10^{-3}$ | 20.4234 |
| $P$ [J m$^{-3}$] | $1.5867 \times 10^2$ | $1.3916 \times 10^6$ |
| $c_V$ [J m$^{-3}$ K$^{-1}$] | $2.2705 \times 10^{-2}$ | 61.2702 |
| $n$ [m$^{-3}$] | $4.2168 \times 10^{19}$ | $4.1073 \times 10^{23}$ |
| $h(T)$ [J m$^{-3}$] | $1.5867 \times 10^2$ | $1.3916 \times 10^6$ |

**Table 2** Calculated values of the radiative and thermodynamic state functions for the Bose-Einstein (μ-type) spectrum and monopole spectrum at $z = 10^5$ and $T = 2.72550 \times 10^5$ K.

| Quantity | $\mu$-distortion spectrum $\mu = 270$ | Monopole spectrum $\mu = 0$ |
|---|---|---|
| $I_0(T)$ [J m$^{-3}$] | $1.7188 \times 10^{-105}$ | $3.3814 \times 10^{12}$ |
| $I_0^{SB}(T)$ [W m$^{-2}$] | $1.2885 \times 10^{-97}$ | $2.5344 \times 10^{20}$ |
| $\Omega(T)$ [J m$^{-3}$] | $-5.7295 \times 10^{-106}$ | - |
| $f$ [J m$^{-3}$] | $-1.5527 \times 10^{-103}$ | $-1.1272 \times 10^{12}$ |
| $s$ [J m$^{-3}$ K$^{-1}$] | $1.9200 \times 10^{-110}$ | $5.5142 \times 10^{5}$ |
| $P$ [J m$^{-3}$] | $5.7295 \times 10^{-106}$ | $1.1272 \times 10^{12}$ |
| $c_V$ [J m$^{-3}$ K$^{-1}$] | $5.7600 \times 10^{-110}$ | $1.6543 \times 10^{6}$ |
| $n$ [m$^{-3}$] | $5.0754 \times 10^{-90}$ | $1.1089 \times 10^{28}$ |
| $h(T)$ [J m$^{-3}$] | $5.7295 \times 10^{-106}$ | $1.1272 \times 10^{12}$ |

**Table 3** Calculated values of the radiative and thermodynamic state functions for the Bose-Einstein (μ-type) spectrum and monopole spectrum at $z = 3 \times 10^6$ and $T = 8.176442 \times 10^6$ K.